\documentclass[preprintnumbers,amsmath,11pt,amssymb,floatfix,superscriptaddress,nofootinbib]{revtex4}
\usepackage{graphicx}
\usepackage{epsfig}
\usepackage{bm}
\usepackage{amsfonts}
\def\be {\begin{equation}}
\def\ee {\end{equation}}
\def\ba {\begin{eqnarray}}
\def\ea {\end{eqnarray}}
\begin{document}

\title{\Large Thermodynamics of Evolving Lorentzian Wormhole at Apparent and Event Horizons}

\author{{\bf Ujjal Debnath}}
\email{ujjaldebnath@yahoo.com , ujjal@iucaa.ernet.in}
\affiliation{Department of Mathematics, Bengal Engineering and
Science University, Shibpur, Howrah-711 103, India.}

\author{\textbf{Mubasher Jamil}}
\email{mjamil@camp.nust.edu.pk} \affiliation{Center for Advanced
Mathematics and Physics, National University of Sciences and
Technology, Islamabad, Pakistan.} \affiliation{Eurasian
International Center for Theoretical Physics, Eurasian National
University, Astana 010008, Kazakhstan.}

\author{\textbf{Ratbay Myrzakulov}}
\email{rmyrzakulov@csufresno.edu, rmyrzakulov@gmail.com}
\affiliation{Eurasian International Center for Theoretical
Physics, Eurasian National University, Astana 010008, Kazakhstan.}

\author{\textbf{M. Akbar}}
\email{makbar@camp.nust.edu.pk} \affiliation{Center for Advanced
Mathematics and Physics, National University of Sciences and
Technology, Islamabad, Pakistan.}

\begin{abstract}\textbf{\\Abstract:}
We have investigated the non-static Lorentzian Wormhole model in
presence of anisotropic pressure. We have presented some exact solutions of Einstein equations for anisotropic pressure case. Introducing two EoS parameters we have shown that these solutions give very rich dynamics of the universe yielding to the different expansion history of it in the $r$ - direction and in the $T$ - direction.  The corresponding explicit forms of the shape function $b(r)$ is presented.We have shown that the
Einstein's field equations and unified first law are equivalent
for the dynamical wormhole model. The first law of thermodynamics has been
derived by using the Unified first law. The physical quantities including surface gravity and the
temperature are derived for the wormhole. Here we have obtained all
the results without any choice of the shape function. The validity
of generalized second law (GSL) of thermodynamics has been
examined at apparent and event horizons for the evolving
Lorentzian wormhole.
\end{abstract}

\maketitle
\newpage

\section{Introduction}

There are a number of similarities between black-hole physics and
thermodynamics. Most striking is the similarity in the behaviors
of black-hole area and of entropy: Both quantities tend to
increase irreversibly. Employing the concepts of information
theory to the black hole physics, Bekenstein introduced the
concept of black-hole entropy as the measure of information about
a black hole interior which is inaccessible to an exterior
observer. Moreover dimensional considerations indicated that the
black-hole entropy is equal to the ratio of the black-hole area to
the square of the Planck length times a dimensionless constant of
order unity \cite{bek}. Later on he deduced the generalized second
law (GSL) of thermodynamics which stated that the combined entropy
of black horizon and `common entropy' does not decrease
\cite{bek1}. Numerous approaches have been utilized to prove the
GSL \cite{sorkin}, while this law has found numerous applications
in cosmology \cite{gsl}. The connection between gravity and
thermodynamics was extended to cosmological horizons with
repulsive cosmological constant \cite{hawk}. Hawking showed
\cite{hawking} that black holes emit thermal radiation
corresponding to a temperature proportional to surface gravity and
entropy proportional to the horizon area $(S\propto A/4)$. This
entropy-area relation was also proved via other approaches in
\cite{app}. Birrel \& Davies also confirmed the thermal nature of
the emitted radiation for the massless Thirring model of a
self-interacting fermion field in a curved two-dimensional
background spacetime \cite{davies}.  The horizon temperature and
entropy obey a simple differential relationship $dE=TdS$, called
the first law of black hole thermodynamics \cite{bardeen}, where
$E$ is the energy. Wald deduced that black hole thermodynamics is
nothing more than ordinary thermodynamics applied to a
self-gravitating quantum system \cite{wald}. Unruh and Wald
described how energy from a black hole can me mined under the
Bekenstein entropy-energy radio \cite{wald1}. Li \& Liu pointed
out that the Unruh-Wald conclusion does not hold because Hawking
radiation near the horizon is not thermal \cite{liu}. Zurek \&
Thorne showed that  entropy of a rotating, charged black hole is
equal to the logarithm of the number of quantum mechanically
distinct ways that the hole could have been made \cite{zurek}.
Visser showed that Hawking radiation can occur in physical
situations in which the laws of black hole mechanics do not apply,
and in physical situations in which the notion of black hole
entropy does not even make any sense \cite{visser}. In recent
years, the phenomenon of Hawking radiation is also studied in the
frameworks of string theory and loop quantum gravity
\cite{strings}. Another significant development was made by
Jacobson \cite{jacobson} by deriving Einstein field equations from
the proportionality of entropy to the horizon area together with
the fundamental relation $\delta Q=TdS$, where $\delta Q$ and $T$
are the energy flux and Unruh temperature seen by an accelerated
observer just inside the horizon.

Padmanabhan \cite{Padmanabhan} made the major development by
launching a general formalism for the spherically symmetric black
hole spacetimes to understand the thermodynamics of horizons and
showed that the Einstein field equations evaluated at event
horizon can be expressed in the form, $TdS = dE + pdV$, of
thermodynamics. Later on Padmanahban et al and others
\cite{paranjape,paranjape1} studied this approach for more general
spacetime geometries and in various gravity theories. In the
cosmological setup, Cai and his collaborators
\cite{cai,cai1,akbar1, cai2, akbar} made the major development by
showing that the Einstein field equations evaluated at the
apparent horizon can also be expressed as $TdS = dE + WdV$ in
various theories of gravity. This connection between gravity and
thermodynamics has also been extended in the braneworld cosmology
\cite{wang}. More recently, using Clausius relation $\delta
Q=TdS$, to the apparent horizon of a FRW universe, Cai et al  are
able to derive the modified Friedman equation by employing quantum
corrected area-entropy formula \cite{rong}.  All these
calculations indicate that the thermal interpretation of gravity
is to be generic, so we have to investigate this relation for a
more general spacetimes.

In this work, we employ the metric of an evolving Lorentzian
wormhole \cite{cata} and aim to show that the Einstein field
equations and the Unified first law are equivalent. We have shown
that the isotropic pressure for non-static wormhole generates the
standard FRW model. The non-static wormhole exits only for
anisotropic pressure. The previous works of Jamil et al \cite{JA},
Farook et al \cite{FA} and Rahaman et al \cite{Ak} have some
computational errors for wormhole thermodynamics in presence of
isotropic pressure. In this work, we have corrected these
assigning with anisotropic pressure in the field equations. To
evaluate the thermodynamical quantities, we use the apparent and
event horizons of the evolving wormhole.

The plan of the paper as follows.
In section II, we write down the field equations and energy
conservation equation for the evolving wormhole. In section III,
we study the wormhole thermodynamics using first law of
thermodynamics and the entropy-area law. The conclusion is
presented at the end of the work.

\section{Evolving Lorentzian Wormhole}

A wormhole consists of a tunnel of trapped surfaces between two
mouths, defined as temporal outer trapping horizons with opposite
senses, in mutual causal contact \cite{morris}. In static cases,
the mouths coincide as the throat of a Morris-Thorne (MT)
wormhole. To keep the wormhole's throat open, an exotic fluid
violating the null energy condition is required \cite{jamil1}. The
zeroth, first and second laws are derived for wormholes are
derived in \cite{hay}.  A simple generalization of Morris-Thorne
wormhole to the time dependent background is given by the evolving
Lorentzian wormhole \cite{cata} \be
ds^2=-e^{2\Phi(t,r)}dt^2+a^2(t)\Big[
\frac{dr^2}{1-\frac{b(r)}{r}}+ r^2d\Omega_2^2 \Big]. \ee Here
$d\Omega_2^2\equiv d\theta^2+\sin^2\theta d\phi^2$ is the line
element of two dimensional unit sphere, $b(r)$ and $\Phi(t,r)$ are
the shape and potential functions respectively and $a(t)$ is the
scale factor of the universe. It is clear from the metric (1) that
if both $b(r)\rightarrow kr^{3}$, and $\Phi(t,r)\rightarrow$
constant, the above metric reduces to the FRW metric. Furthermore,
when $a(t)\rightarrow$ constant and
 $\Phi(t,r)\rightarrow \Phi(r)$, it turns out the static MT wormhole
 \cite{morris}. If one takes $a(t)=e^{\chi t}$, the metric (1) represents
 an inflating Lorentzian wormhole \cite{roman}, where the arbitrary
 constant $\chi$ can be fixed by taking it a cosmological constant
 $\Lambda$. Now consider the components of energy-momentum tensor
 \cite{cata}
 \be
T_{t}^{t}=-\rho(t,r),~T_{r}^{r}=p_{r}(t,r),~T_{\theta}^{\theta}=T_{\phi}^{\phi}=p_{T}(t,r),
 \ee
where $\rho(t,r)$, $p_{r}(t,r)$ and $p_{T}(t,r)$ are the energy
density, radial pressure and tangential pressure respectively. If
$p_{T}=p_{r}$ then the pressure will be isotropic otherwise
anisotropic. From the Einstein's equation $G_{\mu\nu}=8\pi
GT_{\mu\nu}$, we get the following field  equations: \cite{cata}
\be 3e^{-2\Phi}H^{2}+\frac{b'}{a^{2}r^{2}}=8\pi G\rho, \ee \be
-e^{-2\Phi}(2\dot{H}+3H^{2})+2e^{-2\Phi}\dot{\Phi}H-\frac{b}{a^{2}r^{3}}=8\pi
Gp_{r}, \ee \be
-e^{-2\Phi}(2\dot{H}+3H^{2})+2e^{-2\Phi}\dot{\Phi}H+\frac{b-rb'}{2a^{2}r^{3}}=8\pi
Gp_{T}, \ee \be 2\Phi'H=0 ,\ee where $H=\frac{\dot{a}}{a}$ is the
Hubble parameter and dot and dash refer to derivative w.r.t. $t$
and $r$ respectively. Now the equation (6) implies $\Phi'=0$ i.e.,
$\Phi(t,r)=\Phi(t)$ i.e., $\Phi$ is a function of time only. So
without any loss of generality, by rescaling the time coordinate
we can set $\Phi=0$. So the field equations (3) - (5) reduces to
\be 3H^{2}+\frac{b'}{a^{2}r^{2}}=8\pi G\rho, \ee \be
2\dot{H}+3H^{2}+\frac{b}{a^{2}r^{3}}=-8\pi Gp_{r}, \ee \be
2\dot{H}+3H^{2}-\frac{b-rb'}{2a^{2}r^{3}}=-8\pi Gp_{T}. \ee From
equations (8) and (9) we obtain the following form, \be
\frac{rb'-3b}{2a^{2}r^{3}}=8\pi G(p_{r}-p_{T}). \ee Now from the
energy conservation equation $T^{\mu}_{\nu;\mu}=0$, we have \be
\dot{\rho}+H(3\rho+p_{r}+2p_{T})=0, \ee \be
2(p_{T}-p_{r})=rp'_{r}. \ee For isotropic pressure $p_{T}=p_{r}$
and so from (12) we get, $p'_{r}=0$ i.e., $p_{r}$ is function of
time $t$ only. In this case, from (10), we find $b(r)=kr^{3}$ and
hence the metric (1) reduces to FRW metric. In this case (7) and
(12) imply $p_{T}$ and $\rho$ are functions of $t$ only. Since we
want to study the wormhole model with pressure depending on both
variables $t$ and $r$, so we must consider only anisotropic
pressures, thus requiring $p_{r}\ne p_{T}$. One of interesting conseques of  consideration anisotropic pressures is (as we see below) different dynamics of the universe in $r$ - direction and in $T$ - direction. To demonstrate this phenomen let us consider the simple power-law solution:
 $a=a_0t^n$. Then $H=nt^{-1}$.  Substituting these expressions into (7)-(9) we get (below we assume $8\pi G=1$)
\begin{eqnarray}
\rho&=&\frac{3n^2}{t^{-2}}+\frac{b'}{a^{2}_0t^{2n}r^{2}}, \\
p_r&=&\frac{n(2-3n)}{t^2}-\frac{b}{a^{2}_0t^{2n}r^{3}}, \\
 p_T&=&\frac{n(2-3n)}{t^2}+\frac{b-rb'}{2a^{2}_0t^{2n}r^{3}}.
\end{eqnarray}
Now let us introduce separate two  EoS parameters for the  $r$ - direction and for the  $T$ - direction as
\be
\omega_r=\frac{p_r}{\rho},\quad
\omega_T=\frac{p_T}{\rho}.
 \ee
 Then for the solutions (13)-(15) we obtain
 \begin{eqnarray}
\omega_r&=&\frac{n(2-3n)a^{2}_0r^{3}t^{2n-2}-b}{r[3n^2a^{2}_0r^{2}t^{2n-2}+b^{'}]}, \\
\omega_T&=&\frac{2n(2-3n)a^{2}_0r^{3}t^{2n-2}+(b-rb^{'})}{2r[3n^2a^{2}_0r^{2}t^{2n-2}+b^{'}]}.
\end{eqnarray}
In these formulas we have one arbitrary function $b(r)$ and two constant parameters $n$ and $a_0$. To find the explicit form of unknown $b(r)$, as example,  we assume that in the $r$-direction we have  an accelerated expansion so that we can put $\omega_r=-1$. Then from (17) we determine the unknown function $b(r)$  as
\be
b(r)=r[C-na^{2}_0r^{2}t^{2n-2}],
 \ee
 where $C=constant$. To eliminate the dependence of this function of $t$, we put $n=1$. Then finally we get
 \be
b(r)=r[C-a^{2}_0r^{2}].
 \ee
 So for the EoS parameters we get
  \begin{eqnarray}
\omega_r&=&-1, \\
\omega_T&=&0
\end{eqnarray}
that corresponds the accelerated expansion of the universe in $r$ - direction and dust matter dominated case in the $T$ - direction.

ii) Our next example is also the power-law solution but with  $\omega_r=const=\omega_{r0}$. Then for the density of energy, pressures and EoS parameters we get the same  expressions as in the previous case. To find $b(r)$ we use again (17) and get the expression
\be
b(r)=Cr^{-\frac{1}{\omega_{r0}}}+\frac{2n-3(1+\omega_{r0})n^2}{1+3\omega_{r0}}a^{2}_0r^{3}t^{2n-2},
 \ee
 where $C=constant$. To eliminate  the $t$ defendence of $b$ we again  put $n=1$. Then finally we get
 \be
b(r)=Cr^{-\frac{1}{\omega_{r0}}}-a^{2}_0r^{3},
 \ee
 In our case the formulas (17)-(18) become
  \begin{eqnarray}
\omega_r&=&-\frac{a^{2}_0r^{3}+b}{r[3a^{2}_0r^{2}+b^{'}]}, \\
\omega_T&=&\frac{b-rb^{'}-2a^{2}_0r^{3}}{2r[3a^{2}_0r^{2}+b^{'}]}.
\end{eqnarray}
So from these formulas and (24) finally we obtain
  \begin{eqnarray}
\omega_r&=&\omega_{r0}, \\
\omega_T&=&-\frac{1+\omega_{r0}}{2r}.
\end{eqnarray}
Consider particular cases. 1) Let $\omega_{r0}=1/3$ that is raditation. Then
 \begin{eqnarray}
\omega_r&=&1/3, \\
\omega_T&=&-\frac{2}{3r}.
\end{eqnarray}
2) Let $\omega_{r0}<-1$ that is phantom matter. Then
 \begin{eqnarray}
\omega_r&<&-1, \\
\omega_T&>&0.
\end{eqnarray}
This means that in the $r$ - direction we have the radiation dominated dynamics but in the $T$ - direction more complicated one. For $r=r_0=3/2$ we have the transion point from the phantom to the quintessense case.

3) Let $\omega_{r0}>1$ that is ekpyrotic matter. Then
 \begin{eqnarray}
\omega_r&>&1, \\
\omega_T&<&-\frac{1}{r}.
\end{eqnarray}
It is interesting to note that in this case we have the ekpyrotic matter in $r$-direction but  phantom in $T$-direction if $r<1$. So that $r=1$ is a transion point.
\section{Wormhole Thermodynamics}

Thermal properties of wormholes have been studied in the
literature. Hong \& Kim constructed the wormhole's entropy and
Hawking temperature by exploiting Unruh effects and proposed a
possibility of negative temperature originated from exotic matter
distribution of the wormhole \cite{kim}. In \cite{jamil}, the
authors have shown that the Einstein field equations can be
rewritten as a similar form of the first law of thermodynamics at
the dynamical trapping horizon for the (2+1)-dimensional evolving
wormhole spacetime. In \cite{akbar2}, the authors studied the
generalized second law of thermodynamics at the apparent horizon
of the evolving wormhole. In \cite{faiz}, the authors studied the
validity of the generalized second law of thermodynamics by
assuming the logarithmic correction to the horizon entropy of an
evolving wormhole. In \cite{pedro}, the author has shown the
validity of the generalize second law for a Euclidean wormhole.

We quote the laws of wormhole thermodynamics \cite{pedro1}, as our
later use ``First law: The change in the gravitational energy of a
wormhole equals the sum of the energy removed from the wormhole
plus the work done in the wormhole. Second law: The entropy of a
dynamical wormhole is given by its surface area which always
increases. Third law: It is impossible to reach the absolute zero
for surface gravity by any dynamical process.''

We consider the metric in the following form \cite{cai}
\begin{equation}
ds^{2}
=h_{ij}dx^{i}dx^{j}+\tilde{r}^{2}d\Omega_{2}^{2}~~,~~i,j=0,1
\end{equation}
where, $h_{ij}=\left(-1, a^{2}\left(1-\frac{b(r)}{r}\right)^{-1}
\right)$. Now write, $\tilde{r}=ar$. From this we get,
$\dot{\tilde{r}}=\tilde{r}H$. The unified first law is defined by
\cite{hay1}
\begin{equation}
dE=A\Psi+WdV,
\end{equation}
where
\begin{equation}
A=4\pi \tilde{r}^{2},
\end{equation}
is the area and the volume $V$ is defined by
\begin{equation}
V=\frac{4}{3}\pi \tilde{r}^{3}.
\end{equation}

The unified first law (14) expresses the gradient of the active
gravitational energy $E$ according to the Einstein equation,
divided into energy-supply and work terms. The first term on the
right hand side could be interpreted as an energy supply term,
i.e., this term produces a change in the energy of the spacetime
due to the energy flux $\Psi$ generated by the surrounding
material (which generates this geometry). The second term $W$
behaves like a work term, something like the work that the matter
content must do to support this configuration \cite{hay1,pedro1}.

The work density function is given by
\begin{equation}
W=-\frac{1}{2}h^{ij}T_{ij}=\frac{1}{2}~(\rho-p_{r}).
\end{equation}
The energy-supply vector is given by
\begin{equation}
\Psi_{i}=h^{j\lambda}T_{i\lambda}\partial_{j}
(\tilde{r})+W\partial_{i}
(\tilde{r})=\left(-\frac{1}{2}(\rho+p_{r})\tilde{r}H,\frac{1}{2}(\rho+p_{r})a
\right).
\end{equation}
So we have
\begin{equation}
\Psi=\Psi_{i}dx^{i}=\frac{1}{2}(\rho+p_{r})(-\tilde{r}Hdt+adr).
\end{equation}
The energy inside the surface is given by
\begin{equation}
E=\frac{4\pi \tilde{r}}{8\pi
G}\left(1-h^{ij}\partial_{i}\tilde{r}\partial_{j}\tilde{r}\right)=
\frac{\tilde{r}}{2G} \left(\tilde{r}^{2}H^{2}+\frac{b}{r}\right).
\end{equation}
Now we get
\begin{equation}
A\Psi+WdV=-4\pi \tilde{r}^{3}Hp_{r}dt+4\pi a\tilde{r}^{2}\rho dr.
\end{equation}
From (20), we get
\begin{equation}
dE=\frac{\tilde{r}H}{2G}\left[\tilde{r}^{2}(2\dot{H}+3H^{2})+\frac{b}{r}
\right]dt +\frac{1}{2G}\left[3a\tilde{r}^{2}H^{2}+ab' \right]dr.
\end{equation}

Using (21) and (22) and the unified first law (14), on comparing the
coefficients of $dt$ and $dr$, we can directly obtained the field
equations (7) and (8). Also using the conservation equation (11),
the last field equation (9) can be obtained.\\

Now the Gibb's law of thermodynamics states that
\begin{equation}
T_{h}dS_{I}=\frac{1}{3}(p_{r}+2p_{T})dV+d(\rho V),
\end{equation}
where $S_{I}$ is the entropy within the horizon and assume the
average pressure inside the horizon. The variation of internal
entropy is obtained as
\begin{equation}
T_{h}dS_{I}=\frac{4\pi\tilde{r}_{h}^{2}}{3}(3\rho+p_{r}+2p_{T}+\frac{\tilde{r}_{h}\rho'}{a})(d\tilde{r}_
{h}-H\tilde{r}_{h}dt).
\end{equation}

In the following subsection, the first law of thermodynamics will
be derived using unified first law. Then the GSL will be examined
for apparent and event horizons of the wormhole using first law of
thermodynamics.

\subsection{\large Using First Law (of Thermodynamics)}

We know that heat is one of the form of energy. Therefore, the
heat flow $\delta Q$ through the horizon is just the amount of
energy crossing it during the time interval $dt$. That is, $\delta
Q=-dE$ is the change of the energy inside the horizon. So from
equation (14) and (21) we have the amount of the energy crossing
on the horizon as
\begin{equation}
-dE_{h}=4\pi \tilde{r}_{h}^{3}Hp_{r}dt-4\pi \tilde{r}_{h}^{2}\rho
(d\tilde{r}_{h}-H\tilde{r}_{h}dt)=4\pi
\tilde{r}_{h}^{3}H(\rho+p_{r})dt-4\pi \tilde{r}_{h}^{2}\rho
d\tilde{r}_{h}.
\end{equation}
From this, we see that there is no effect of density and
tangential pressure on the horizon. The first law of
thermodynamics (Clausius relation) on the horizon is defined as
follows:
\begin{equation}
T_{h}dS_{h}=dQ=-dE_{h}.
\end{equation}
From these equations, the variation of entropy on the horizon is
given by
\begin{equation}
T_{h}dS_{h}=4\pi \tilde{r}_{h}^{3}H(\rho+p_{r})dt-4\pi
\tilde{r}_{h}^{2}\rho d\tilde{r}_{h}.
\end{equation}
From (24) and (27), we obtain the variation of total entropy as
\begin{equation}
T_{h}\dot{S}_{total}=\frac{4\pi\tilde{r}_{h}^{2}}{3}(p_{r}+2p_{T}+\frac{\tilde{r}_{h}\rho'}{a})\dot{\tilde{r}}_{h}
+\frac{8\pi\tilde{r}_{h}^{3}H}{3}(2p_{r}-2p_{T}-\frac{\tilde{r}_{h}\rho'}{a}),
\end{equation}
which becomes
\begin{eqnarray*}
T_{h}\dot{S}_{total}=\frac{1}{6G}\left[-3\tilde{r}_{h}^{2}(2\dot{H}+3H^{2})-3b'(\tilde{r}_{h}/a)
+\frac{b''(\tilde{r}_{h}/a)}{a\tilde{r}_{h}}
\right]\dot{\tilde{r}}_{h}~~~~~~~~~~~~~~~~~~~~~~~~
\end{eqnarray*}
\begin{equation}
+\frac{H}{6G}\left[\left\{\tilde{r}b(\tilde{r}_{h}/a)-3ab(\tilde{r}_{h}/a)\right\}-\frac{2\tilde{r}_{h}}{a}
\left\{\tilde{r}_{h}b''(\tilde{r}_{h}/a)-2ab'(\tilde{r}_{h}/a)
\right\} \right].
\end{equation}

Now we shall analyze the apparent and event horizons for wormhole
and find out the radius on both horizons and investigate the GSL
of thermodynamics in general way.

\subsubsection{\bf Apparent Horizon}

The dynamical apparent horizon $\tilde{r}_{A}$, a marginally
trapped surface with vanishing expansion, is determined by the
relation
\begin{equation}
\left[h^{ij}\partial_{i}\tilde{r}\partial_{j}\tilde{r}\right]_{\tilde{r}=\tilde{r}_{A}}=0,
\end{equation}
i.e.,
\begin{equation}
H^{2}\tilde{r}_{A}^{2}=1-\frac{ab(\tilde{r}_{A}/a)}{\tilde{r}_{A}}.
\end{equation}
Taking derivative, we obtain,
\begin{equation}
\dot{\tilde{r}}_{A}=\frac{H\tilde{r}_{A}\{\tilde{r}_{A}b'(\tilde{r}_{A}/a)-ab(\tilde{r}_{A}/a)-2\dot{H}\tilde{r}_{A}^{3}
\}}{\{\tilde{r}_{A}b'(\tilde{r}_{A}/a)-ab(\tilde{r}_{A}/a)+2H^{2}\tilde{r}_{A}^{3}.
\}}
\end{equation}

From (29), we obtain the rate of change of total entropy for
apparent horizon as
\begin{eqnarray*}
T_{A}\dot{S}_{total}=\frac{H\tilde{r}_{A}\{\tilde{r}_{A}b'(\tilde{r}_{A}/a)-ab(\tilde{r}_{A}/a)-2\dot{H}\tilde{r}_{A}^{3}
\}}{6G\{\tilde{r}_{A}b'(\tilde{r}_{A}/a)-ab(\tilde{r}_{A}/a)+2H^{2}\tilde{r}_{A}^{3}
\}}\left[-3\tilde{r}_{A}^{2}(2\dot{H}+3H^{2})-3b'(\tilde{r}_{A}/a)
+\frac{b''(\tilde{r}_{A}/a)}{a\tilde{r}_{A}} \right]
\end{eqnarray*}
\begin{equation}
+\frac{H}{6G}\left[\left\{\tilde{r}b(\tilde{r}_{A}/a)-3ab(\tilde{r}_{A}/a)\right\}-\frac{2\tilde{r}_{A}}{a}
\left\{\tilde{r}_{A}b''(\tilde{r}_{A}/a)-2ab'(\tilde{r}_{A}/a)
\right\} \right].
\end{equation}

The GSL for the apparent horizon will be satisfied if the r.h.s of
the above expression is non-negative.

\subsubsection{\bf Event Horizon}

Event horizon radius $\tilde{r}_{E}$ can be found from the
relation (i.e., $ds^{2}=0=d\Omega_{2}^{2}$)
\begin{equation}
\dot{\tilde{r}}_{E}=\tilde{r}_{E}H-\sqrt{1-\frac{ab(\tilde{r}_{E}/a)}{\tilde{r}_{E}}},
\end{equation}
or
\begin{equation}
\int_{0}^{\frac{\tilde{r}_{E}}{a}}\frac{dr}{\sqrt{1-\frac{b(r)}{r}}}=\int_{t}^{\infty}\frac{dt}{a}.
\end{equation}
From (29), we obtain the rate of change of total entropy for event
horizon as
\begin{eqnarray*}
T_{E}\dot{S}_{total}=\frac{1}{6G}\left[-3\tilde{r}_{E}^{2}(2\dot{H}+3H^{2})-3b'(\tilde{r}_{E}/a)
+\frac{b''(\tilde{r}_{E}/a)}{a\tilde{r}_{E}}
\right]\left(\tilde{r}_{E}H-\sqrt{1-\frac{ab(\tilde{r}_{E}/a)}{\tilde{r}_{E}}}\right)
\end{eqnarray*}
\begin{equation}
+\frac{H}{6G}\left[\left\{\tilde{r}b(\tilde{r}_{E}/a)-3ab(\tilde{r}_{E}/a)\right\}-\frac{2\tilde{r}_{E}}{a}
\left\{\tilde{r}_{E}b''(\tilde{r}_{E}/a)-2ab'(\tilde{r}_{E}/a)
\right\} \right].
\end{equation}

If the above expression is non-negative, we can say that the GSL
is valid for event horizon.\\

In the following subsection, we shall consider the area law of
thermodynamics i.e., the entropy on the horizon is proportional to
the area of the spherical horizon surface. Then the GSL will be
examined for apparent and event horizons of the wormhole using
area law of thermodynamics.

\subsection{\large Using Area Law (of Thermodynamics)}

Now we shall analyze the apparent and event horizons for wormhole
and find out the radius on both horizons and investigate the GSL
of thermodynamics in general way.

\subsubsection{\bf Apparent Horizon}

The surface gravity is defined as
\begin{equation}
\kappa=\frac{1}{2\sqrt{-h}}~\partial_{i}(\sqrt{-h}~h^{ij}\partial_{j}\tilde{r}).
\end{equation}
Here $h=det(h_{ij})$. The dynamical apparent horizon radius
$\tilde{r}_{A}$ is given in equation (31). So we get the surface
gravity on the apparent horizon:
\begin{equation}
\kappa=-\frac{1}{2}~\tilde{r}_{A}(\dot{H}+2H^{2})+\frac{1}{4\tilde{r}_{A}^{2}}
\left[ ab(\tilde{r}_{A}/a)- \tilde{r}_{A}b'(\tilde{r}_{A}/a)
\right].
\end{equation}
Now the apparent horizon temperature is
\begin{equation}
T_{A}=\frac{\kappa}{2\pi}=-\frac{1}{4\pi}~\tilde{r}_{A}(\dot{H}+2H^{2})+\frac{1}{8\pi
\tilde{r}_{A}^{2}} \left[ ab(\tilde{r}_{A}/a)-
\tilde{r}_{A}b'(\tilde{r}_{A}/a) \right].
\end{equation}
Since the area of the wormhole horizon is $A=4\pi
\tilde{r}_{A}^{2}$, so one can relate the entropy with the surface
area of the apparent horizon (area law) through $S_{A} = A/4G$.
Therefore we have
\begin{equation}
S_{A}=\frac{\pi \tilde{r}_{A}^{2}}{G},
\end{equation}
so that
\begin{equation}
dS_{A}=\frac{2\pi \tilde{r}_{A}d\tilde{r}_{A}}{G}.
\end{equation}
Using (24), (39) and (41) we have
\begin{equation}
T_{A}\dot{S}_{total}=T_{A}(\dot{S}_{I}+\dot{S}_{A})=\frac{4\pi\tilde{r}_{A}^{2}}{3}(3\rho+p_{r}+2p_{T}
+\frac{\tilde{r}_{A}\rho'}{a})(\dot{\tilde{r}}_{A}-H\tilde{r}_{A})+\frac{2\pi
\tilde{r}_{A}T_{A}\dot{\tilde{r}}_{A}}{G},
\end{equation}
which can be written as
\begin{eqnarray*}
T_{A}\dot{S}_{total}=\frac{H(H^{2}+\dot{H})\tilde{r}_{A}^{4}
}{3G\{\tilde{r}_{A}b'(\tilde{r}_{A}/a)-ab(\tilde{r}_{A}/a)+2H^{2}\tilde{r}_{A}^{3}
\}}\left[3\dot{H}\tilde{r}_{A}^{2}
-\frac{b''(\tilde{r}_{A}/a)}{a\tilde{r}_{A}}
\right]~~~~~~~~~~~~~~~~~~~~~
\end{eqnarray*}
\begin{equation}
+\frac{H}{2G} \left[ab(\tilde{r}_{A}/a)-
\tilde{r}_{A}b'(\tilde{r}_{A}/a)-2\tilde{r}_{A}^{3}(\dot{H}+2H^{2})
\right]\frac{\{\tilde{r}_{A}b'(\tilde{r}_{A}/a)-ab(\tilde{r}_{A}/a)-2\dot{H}\tilde{r}_{A}^{3}
\}}{\{\tilde{r}_{A}b'(\tilde{r}_{A}/a)-ab(\tilde{r}_{A}/a)+2H^{2}\tilde{r}_{A}^{3}
\}}.
\end{equation}

The GSL for the apparent horizon will be satisfied if the r.h.s of
the above expression is non-negative.

\subsubsection{\bf Event Horizon}

We obtain the rate of change of total entropy for event horizon as
\begin{equation}
T_{E}\dot{S}_{total}=T_{E}(\dot{S}_{I}+\dot{S}_{E})=\frac{4\pi\tilde{r}_{E}^{2}}{3}(3\rho+p_{r}+2p_{T}
+\frac{\tilde{r}_{E}\rho'}{a})(\dot{\tilde{r}}_{E}-H\tilde{r}_{E})+\frac{2\pi
\tilde{r}_{E}T_{E}\dot{\tilde{r}}_{E}}{G},
\end{equation}
which can be written as
\begin{eqnarray*}
T_{E}\dot{S}_{total}=\frac{1}{6G}\left[6\tilde{r}_{E}^{2}\dot{H}
-\frac{b''(\tilde{r}_{E}/a)}{a\tilde{r}_{E}}
\right]\sqrt{1-\frac{ab(\tilde{r}_{E}/a)}{\tilde{r}_{E}}}~~~~~~~~~~~~~~~~~~~~~~~~~~~~~~~~
\end{eqnarray*}
\begin{equation}
+\frac{H}{2G} \left[ab(\tilde{r}_{E}/a)-
\tilde{r}_{E}b'(\tilde{r}_{E}/a)-2\tilde{r}_{E}^{3}(\dot{H}+2H^{2})
\right]\left[\tilde{r}_{E}H-\sqrt{1-\frac{ab(\tilde{r}_{E}/a)}{\tilde{r}_{E}}}
\right].
\end{equation}

If the above expression is non-negative, we can say that the GSL
is valid for event horizon.\\

\section{Conclusions}

We have studied the time-dependent Lorentzian Wormhole model in
presence of anisotropic (i.e., radial and tangential) pressure.
The density and pressure are considered in both $t$ and $r$
dependent. For isotropic pressure, the radial pressure transforms
to function of time only. In this case, we obtain the shape
function in the form $b(r)=kr^{3}$, so the the model reduces to
the standard FRW model. We have shown that the Einstein's field
equations and unified first law are equivalent for the dynamical
wormhole model. We have presented some exact solutions of Einstein equations for anisotropic pressure case. Introducing two EoS parameters we have shown that these solutions give very rich dynamics of the universe yielding to the different expansion history of it in the $r$ - direction and in the $T$ - direction.  The corresponding explicit forms of the shape function $b(r)$ is presented. The first law of thermodynamics has been derived
by using the Unified first law in presence of anisotropic
pressure. The physical quantities including surface gravity
($\kappa$) and the equilibrium temperature ($T$) are derived for
the wormhole model. Here we have obtained all the results like
entropy on the horizons, variation of internal and horizon
entropies in general way without any choice of the shape function.
Finally, the validity of generalized second law (GSL) of
thermodynamics has been examined at apparent and event horizons by
considering first law and area law of thermodynamics for the
evolving Lorentzian wormhole.

\subsection*{Acknowledgment}\small

One of the author (UD) is thankful to IUCAA, Pune, India for warm
hospitality where part of the work was carried out. M. Jamil would
like to thank the Abdus Salam International Center for Theoretical
Physics (ICTP) and Eurasian National University for their warm
hospitalies during which part of this work was completed.

\end{document}